\documentclass[11pt]{elsart}
\usepackage{amsfonts}
\usepackage{slashbox}
\usepackage{amssymb,amsmath} 
\usepackage{mathrsfs} 
\usepackage{graphicx}

\begin{document}

\begin{frontmatter}
\title{A note on the definition of Bayesian Nash equilibrium of a mechanism when strategies of agents are costly actions}
\author{Haoyang Wu\corauthref{cor}}
\corauth[cor]{Corresponding author.} \ead{18621753457@163.com}
\address{Wan-Dou-Miao Research Lab, Room 301, Building 3, 718 WuYi Road,\\
 200051, Shanghai, China.}

\begin{abstract}
In mechanism design theory, a designer would like to implement a desired social choice function which specifies her favorite outcome for each possible profile of agents' types. To do so, the designer constructs a mechanism which describes each agent's feasible strategy set and the outcome function. Generally speaking, each agent's strategy in a mechanism has two possible formats: an action, or a message. In this paper, we focus on the former format and claim that the notion of Bayesian Nash equilibrium of a mechanism should be based on a profit function instead of the conventional utility function when strategies of agents are costly actions. Next, we derive the main result: Given a social choice function which can be implemented by an indirect mechanism in Bayesian Nash equilibrium, if all strategies of agents are costly actions, then it cannot be inferred that there exists a direct mechanism that can truthfully implement the social choice function in Bayesian Nash equilibrium.
\end{abstract}

\begin{keyword}
Bayesian Nash Equilibrium; Mechanism design; Revelation Principle.
\end{keyword}
\end{frontmatter}

\section{Introduction}
In the framework of mechanism design theory \cite{MWG1995, Myerson1979, Myerson1982, Narahari2009}, there are one designer and some agents.\footnote{The designer is denoted as ``She'', and the agent is denoted as ``He''.} Suppose that the designer would like to implement a desired social choice function which specifies her favorite outcome for each possible profile of agents' types. However, each agent's type is modelled as his private property and unknown to the designer. In order to implement a social choice function in Bayesian Nash equilibrium, the designer constructs a mechanism which specifies each agent's feasible strategy set ($i.e.$, the allowed actions of each agent) and an outcome function ($i.e.$, a rule for how agents' actions get turned into a social choice).

Generally speaking, each agent's strategy in a mechanism has two possible formats: an action, or a message (\emph{i.e.}, a plan of action) (see MWG's Book, Page 883, Line 8, \cite{MWG1995}). The distinction between the two formats is that: the former format of strategy is \emph{a real action} which naturally requires some action cost to be performed realistically, whereas the latter format of strategy is \emph{a message of action plan} which is reported by each agent to the designer and hence doesn't need action cost to be performed realistically. In this paper, we focus on the former format of strategy, and investigate the notion of Bayesian Nash equilibrium of a mechanism.

The paper is organized as follows. First, we introduce a notion of profit function (\emph{i.e.}, Definition 1),  and then claim that the definition of Bayesian Nash equilibrium of a mechanism should be based on the profit function instead of the conventional utility function when strategies of agents are costly actions (\emph{i.e.}, Definition 2). Next, we derive the main result (\emph{i.e.}, Proposition 1): Given a social choice function $f$ which is implemented by an indirect mechanism in Bayesian Nash equilibrium, if all strategies of agents in the mechanism are costly actions, then it cannot be inferred that there exists a direct mechanism that can truthfully implement $f$ in Bayesian Nash equilibrium. Section 3 concludes the paper.

\section{Theoretical Analysis}
Consider a setting with one designer and $I$ agents indexed by $i=1,\cdots,I$. Each agent $i$ privately observes his \emph{type} $\theta_{i}$ that determines his preference over elements in an outcome set $X$. The set of possible types for agent $i$ is denoted as $\Theta_{i}$. The vector of agents' types $\theta=(\theta_{1},\cdots,\theta_{I})$ is drawn from set $\Theta=(\Theta_{1},\cdots,\Theta_{I})$ according to probability density $\phi(\cdot)$, and each agent $i$'s \emph{utility function} over the outcome $x\in X$ given his type $\theta_{i}$ is a two-parameter function $u_{i}(x,\theta_{i})$.

A \emph{mechanism} $\Gamma=(S_{1},\cdots,S_{I},g(\cdot))$ is a collection of $I$ strategy sets $S_{1},\cdots,S_{I}$ and an outcome function $g:S_{1}\times\cdots\times S_{I}\rightarrow X$. The mechanism combined with possible types $(\Theta_{1},\cdots,\Theta_{I})$, the probability density $\phi(\cdot)$ over the possible realizations of $\theta\in\Theta_{1}\times\cdots\times\Theta_{I}$, and utility functions $(u_{1}, \cdots, u_{I})$ defines a Bayesian game of incomplete information. The strategy function of each agent $i$ in the game induced by $\Gamma$ is a private function $s_{i}(\cdot): \Theta_{i}\rightarrow S_{i}$. Each strategy set $S_{i}$ contains agent $i$'s possible strategies ($i.e.$, \emph{actions}, or \emph{plans of action}). The outcome function $g(\cdot)$ describes the rule for how agents' strategies get turned into a social choice. A \emph{social choice function} (SCF) is a function $f:\Theta_{1}\times\cdots\times\Theta_{I}\rightarrow X$
that, for each possible profile of the agents' types $\theta_{1},\cdots,\theta_{I}$, assigns a collective choice $f(\theta_{1},\cdots,\theta_{I})\in X$.

\textbf{Note 1:}
As shown above, there are two possible formats of each agent's strategy in the mechanism $\Gamma=(S_{1},\cdots,S_{I},g(\cdot))$: an action, or a message. \\
1) If each agent's strategy is an action, then it should be performed realistically. Hence, it is reasonable to assume that each agent shall spend some action cost (or make some effort which can be quantified as some action cost) to perform his strategy.\\
2) If each agent's strategy is a message, then it is not a real action and hence doesn't need action cost to be performed realistically. $\Box$

In the following discussions, we will focus on the former format of strategy and investigate the notion of Bayesian Nash equilibrium of a mechanism. To simplify representations, we assume that each agent's action cost is only relevant to his strategy and private type, and is independent of the game outcome. \footnote{This assumption can be relaxed without changing the following results.}

\textbf{Definition 1:}
Given a social choice function $f$, consider a mechanism $\Gamma=(S_{1},\cdots,S_{I},g(\cdot))$ that implements it in Bayesian Nash equilibrium.  For each agent $i$ with private type $\theta_{i}$, if his strategy $s_{i}(\theta_{i}): \Theta_{i}\rightarrow S_{i}$ in the game induced by $\Gamma$ is a costly action, then the corresponding action cost is defined by a \emph{cost function} $c_{i}(s_{i}, \theta_{i}): S_{i}\times\Theta_{i}\rightarrow \mathcal{R}^{+}$,  \emph{i.e.}, $c_{i}(s_{i}, \theta_{i})> 0$ for each $s_{i}\in S_{i}$, $\theta_{i}\in\Theta_{i}$. Suppose the outcome yielded by $\Gamma$ is denoted as $x=g(s_{1}, \cdots, s_{I})\in X$ and agent $i$'s utility is denoted by a two-parameter function $u_{i}(x,\theta_{i}): X\times\Theta_{i}\rightarrow\mathcal{R}$, then each agent $i$'s profit is defined by a three-parameter \emph{profit function} $p_{i}(x, s_{i}, \theta_{i}): X\times S_{i}\times\Theta_{i}\rightarrow\mathcal{R}$, and
\begin{equation}
p_{i}(x, s_{i}, \theta_{i}) = u_{i}(x,\theta_{i}) - c_{i}(s_{i}, \theta_{i}). \label{profit function} 
\end{equation}

\textbf{Discussion 1:}
Someone may argue that when each agent's strategy is a costly action, then his utility has already included the action cost. Thus, it is not necessary to introduce another notion of profit function to make confusion.

\textbf{Answer 1:}
Generally speaking, there are two versions of utility functions which are commonly used in the literature of game theory and mechanism design:\\
1) \emph{Version 1}: The utility function of an agent has a parameter which corresponds to the agent's strategy. For example, in Section 13.C of MWG's book (Page 450, the fourth line from the bottom, \cite{MWG1995}), the authors use a three-parameter function $u(w, e|\theta) = w - c(e, \theta)$ to denote the utility of a type $\theta$ agent who plays the strategy $e$ (\emph{i.e.}, the education level) and receives an outcome $w$ (\emph{i.e.}, the wage), where $c(e, \theta)$ denotes the agent's cost of obtaining education level $e$. \footnote{Another example can be seen in Section 14.B (Page 480, Line 1, \cite{MWG1995}). The authors use a two-parameter function $u(w,e) = v(w) - g(e)$ to denote the utility of agent who chooses the strategy $e$ (\emph{i.e.}, the education level) and receive the outcome $w$ (\emph{i.e.}, the wage). Here in the definition of utility function $u(w,e)$, the parameter of agent's type is omitted, and $g(e)$ denotes agent's cost with respect to $e$.} Obviously, \emph{the utility function which has a parameter of agent's strategy has already included agent's cost}, and indeed is equivalent to agent's profit function.

2) \emph{Version 2}: The utility function of an agent does not have any parameter which corresponds to the agent's strategy. For example, in Section 23.B of MWG's book (Page 858, the fifth line from the bottom, \cite{MWG1995}), the authors use a two-parameter function $u_{i}(x,\theta_{i})$ to denote the utility of agent $i$ with type $\theta_{i}$ after obtaining an outcome $x\in X$. Obviously, the utility function without having a parameter of agent's strategy only describes the agent's utility with respect to the outcome, and does not include the action cost spent by the agent to obtain the outcome. Hence, when each agent's strategy is a costly action, the second version of utility function should be replaced by the profit function to \emph{exactly} describe how much each agent benefits from the game induced by a mechanism.  $\Box$

Now let us consider the notion of Bayesian Nash equilibrium of a mechanism. Conventionally, a strategy profile $s^{*}(\cdot)=(s^{*}_{1}(\cdot),\cdots,s^{*}_{I}(\cdot))$ is a \emph{Bayesian Nash equilibrium} of mechanism $\Gamma=(S_{1},\cdots,S_{I},g(\cdot))$ if, for all $i$ and all $\theta_{i}\in\Theta_{i}$, $\hat{s}_{i}\in S_{i}$, there exists
\begin{equation}
  E_{\theta_{-i}}[u_{i}(g(s^{*}_{i}(\theta_{i}),s^{*}_{-i}(\theta_{-i})),\theta_{i})|\theta_{i}]
  \geq
  E_{\theta_{-i}}[u_{i}(g(\hat{s}_{i},s^{*}_{-i}(\theta_{-i})),\theta_{i})|\theta_{i}].\quad\quad\label{old BNE} 
\end{equation}

\textbf{Note 2:}
As shown above, the conventional notion of Bayesian Nash equilibrium of a mechanism is based on a two-parameter utility function $u_{i}(x,\theta_{i}): X\times\Theta_{i}\rightarrow\mathcal{R}$, \emph{i.e.}, the second version of utility function. It should be emphasized that the strategies $s^{*}_{i}, s^{*}_{-i}, \hat{s}_{i}$ appeared in formula (\ref{old BNE}) are only used to compute an outcome $x=g(\cdot)\in X$, and do not act as independent parameters of agent $i$'s utility function. When each agent $i$'s strategy $s_{i}(\theta_{i})$ in the mechanism $\Gamma$ is a costly action, $i.e.$, $c_{i}(s_{i}, \theta_{i})> 0$, then as pointed out in Answer 1, the two-parameter utility function $u_{i}(x, \theta_{i})$ cannot describe his profit. \emph{Since it is the profit that each rational agent really concerns in a game, the profit function should be introduced in the definition of the Bayesian Nash equilibrium of a mechanism}.\footnote{In some limited cases, each agent's action cost may be neglected. By Eq (\ref{profit function}) the two-parameter utility function $u_{i}(x, \theta_{i})$ is equivalent to the three-parameter profit function $p_{i}(x, s_{i}, \theta_{i})$. Therefore, the conventional definition of Bayesian Nash equilibrium of a mechanism based on the second version of utility function holds only in these limited cases.}

\textbf{Definition 2}: The strategy profile
$s^{*}(\cdot)=(s^{*}_{1}(\cdot),\cdots,s^{*}_{I}(\cdot))$ is a
\emph{Bayesian Nash equilibrium} of mechanism
$\Gamma=(S_{1},\cdots,S_{I},g(\cdot))$ if, for all $i$ and all
$\theta_{i}\in\Theta_{i}$, there exists
\begin{equation}
  E_{\theta_{-i}}[p_{i}(g(s^{*}_{i}(\theta_{i}),s^{*}_{-i}(\theta_{-i})), s^{*}_{i}(\theta_{i}), \theta_{i})|\theta_{i}] 
  \geq
  E_{\theta_{-i}}[p_{i}(g(\hat{s}_{i}, s^{*}_{-i}(\theta_{-i})), \hat{s}_{i}, \theta_{i})|\theta_{i}]\label{new BNE}
\end{equation}
$i.e.$,
\begin{align*}
  E_{\theta_{-i}}&[(u_{i}(g(s^{*}_{i}(\theta_{i}),s^{*}_{-i}(\theta_{-i})), \theta_{i}) - c_{i}(s^{*}_{i}(\theta_{i}), \theta_{i}))|\theta_{i}]
  \geq\\
  &E_{\theta_{-i}}[(u_{i}(g(\hat{s}_{i},s^{*}_{-i}(\theta_{-i})), \theta_{i}) - c_{i}(\hat{s}_{i}, \theta_{i}))|\theta_{i}],
\end{align*}
for all $\hat{s}_{i}\in S_{i}$, in which $p_{i}$ is the profit of agent $i$ given by Eq (\ref{profit function}).

According to MWG book \cite{MWG1995}, the mechanism
$\Gamma=(S_{1},\cdots,S_{I},g(\cdot))$ \emph{implements the social choice
function} $f(\cdot)$ \emph{in Bayesian Nash equilibrium} if there is a
Bayesian Nash equilibrium of $\Gamma$,
$s^{*}(\cdot)=(s^{*}_{1}(\cdot),\cdots,s^{*}_{I}(\cdot))$, such that
$g(s^{*}(\theta))=f(\theta)$ for all $\theta\in\Theta$. A \emph{direct mechanism} is a mechanism $\bar{\Gamma}=(\bar{S}_{1}, \cdots, \bar{S}_{I}$, $\bar{g}(\cdot))$ in which $\bar{S}_{i}=\Theta_{i}$ for all $i$ and $\bar{g}(\theta)=f(\theta)$ for all $\theta\in\Theta_{1}\times\cdots\times\Theta_{I}$. \footnote{Here we use a bar symbol to distinguish a direct mechanism from an indirect mechanism.}
The social choice function $f(\cdot)$ is
\emph{truthfully implementable in Bayesian Nash equilibrium} (or \emph{Bayesian incentive compatible}) if
$\bar{s}^{*}_{i}(\theta_{i})=\theta_{i}$ for all
$\theta_{i}\in\Theta_{i}$ and $i=1,\cdots,I$ is a Bayesian Nash
equilibrium of the direct mechanism $\Bar{\Gamma}=(\bar{S}_{1},\cdots,\bar{S}_{I},\bar{g}(\cdot))$,
in which $\bar{S}_{i}=\Theta_{i}$, $\bar{g}=f$. That is, if for
all $i=1,\cdots,I$ and all $\theta_{i}\in\Theta_{i}$, $\hat{\theta}_{i}\in \Theta_{i}$, there exists
\begin{equation}
  E_{\theta_{-i}}[u_{i}(f(\theta_{i},\theta_{-i}),\theta_{i})|\theta_{i}]
  \geq
  E_{\theta_{-i}}[u_{i}(f(\hat{\theta}_{i},\theta_{-i}),\theta_{i})|\theta_{i}].\quad\quad\label{old BIC} 
\end{equation}

\textbf{Note 3:}
In the direct mechanism $\bar{\Gamma}=(\bar{S}_{1}, \cdots, \bar{S}_{I}$, $\bar{g}(\cdot))$, each agent $i$ independently chooses his report strategy $\bar{s}_{i}(\cdot): \Theta_{i}\rightarrow\Theta_{i}$, and the report type $\bar{s}_{i}(\theta_{i})$ does not need to be his true type $\theta_{i}$. Hence, the format of each agent $i$'s strategy is not a real action but a message. By Note 1, it is reasonable to assume each agent $i$ plays his strategy costlessly.\footnote{Some researchers investigated misreporting costs in a direct mechanism \cite{Kartik2009, Kephart2016}, which are possibly spent by agents when reporting a false type. However, the misreporting cost is irrelevant to this paper. Our result holds no matter whether there exists the misreporting cost or not. Hence, we simply omit the misreporting cost in this paper.} Thus, by Eq (\ref{profit function}), each agent's utility in the direct mechanism is just equal to his profit. Consequently, although the notion of Bayesian Nash equilibrium of a mechanism should be revised to Definition 2 when strategies of agents are costly actions, the conventional notion of Bayesian incentive compatibility still holds as specified by formula (\ref{old BIC}).

\textbf{Note 4:}
In the direct mechanism $\bar{\Gamma}=(\bar{S}_{1}, \cdots, \bar{S}_{I}$, $\bar{g}(\cdot))$, the only thing that the designer gets from each agent $i$ is the report type $\bar{s}_{i}\in\Theta_{i}$, and she has no way to verify whether these reports are truthful or not. All that the designer can do is to announce $f(\bar{s}_{1}, \cdots, \bar{s}_{I})$ as the outcome. Thus, \emph{in the direct mechanism $\bar{\Gamma}$, each agent i with type $\theta_{i}$ does not need to perform any other strategy $s_{i}(\theta_{i})\in S_{i}$ specified in any indirect mechanism $\Gamma=(S_{1},\cdots,S_{I},g(\cdot))$}, and consequently does not need to spend any action cost. \footnote{Someone may argue that in a direct mechanism $\bar{\Gamma}=(\bar{S}_{1}, \cdots, \bar{S}_{I}$, $\bar{g}(\cdot))$, in addition to choose a type $\bar{s}_{i}\in\Theta_{i}$ to report, each agent with type $\theta_{i}$ may also be \emph{willing} to perform another action $s_{i}(\theta_{i})$ voluntarily as what he would perform in some indirect mechanism $\Gamma=(S_{1},\cdots,S_{I},g(\cdot))$. Thus, each agent $i$ also spends action cost $c_{i}(s_{i}, \theta_{i})$ in the direct mechanism $\bar{\Gamma}$. However, this argument requires each agent to do beyond the framework of the direct mechanism, since the additional strategy $s_{i}(\theta_{i})\in S_{i}$ is meaningless and not defined in the direct mechanism $\bar{\Gamma}$.}

\textbf{Proposition 1:} For a given social choice function $f$, suppose that there exists an indirect mechanism $\Gamma=(S_{1},\cdots,S_{I},g(\cdot))$ that implements it in Bayesian Nash equilibrium. For each agent with type $\theta_{i}$, if his strategy $s_{i}\in S_{i}$ is a costly action, $i.e.$, $c_{i}(s_{i}, \theta_{i})> 0$, then it cannot be inferred that there exists a direct mechanism that can truthfully implement $f$ in Bayesian Nash equilibrium.

\textbf{Proof:}
  Consider the social choice function $f$, and the indirect mechanism $\Gamma=(S_{1},\cdots,S_{I},g(\cdot))$ that implements it in Bayesian Nash equilibrium, then there exists a profile of strategies $s^{*}(\cdot)=(s^{*}_{1}(\cdot),\cdots,s^{*}_{I}(\cdot))$ such that the mapping $g(s^{*}(\cdot)):\Theta_{1}\times\cdots\times\Theta_{I}\rightarrow X$ from a vector of agents' types $
\theta=(\theta_{1},\cdots,\theta_{I})$ into an outcome $g(s^{*}(\theta))$ is equal to the desired outcome $f(\theta)$, $i.e.$, $g(s^{*}(\theta))=f(\theta)$ for all $\theta\in\Theta_{1}\times\cdots\times\Theta_{I}$.

By Definition 2, for all $i$ and all $\theta_{i}\in\Theta_{i}$, $\hat{s}_{i}\in S_{i}$,
\begin{align*}
  E_{\theta_{-i}}&[(u_{i}(g(s^{*}_{i}(\theta_{i}),s^{*}_{-i}(\theta_{-i})), \theta_{i}) - c_{i}(s^{*}_{i}(\theta_{i}), \theta_{i}))|\theta_{i}]
  \geq\\
  &E_{\theta_{-i}}[(u_{i}(g(\hat{s}_{i},s^{*}_{-i}(\theta_{-i})), \theta_{i}) - c_{i}(\hat{s}_{i}, \theta_{i}))|\theta_{i}].
\end{align*}
Thus, for all $i$ and all $\theta_{i}\in\Theta_{i}$, $\hat{\theta}_{i}\in \Theta_{i}$,
\begin{align*}
  E_{\theta_{-i}}&[(u_{i}(g(s^{*}_{i}(\theta_{i}),s^{*}_{-i}(\theta_{-i})), \theta_{i}) - c_{i}(s^{*}_{i}(\theta_{i}), \theta_{i}))|\theta_{i}]
  \geq\\
  &E_{\theta_{-i}}[(u_{i}(g(s^{*}_{i}(\hat{\theta}_{i}),s^{*}_{-i}(\theta_{-i})), \theta_{i}) - c_{i}(s^{*}_{i}(\hat{\theta}_{i}), \theta_{i}))|\theta_{i}].
\end{align*}

Since $g(s^{*}(\theta))=f(\theta)$ for all $\theta$, then for all $i$ and all $\theta_{i}\in\Theta_{i}$, $\hat{\theta}_{i}\in \Theta_{i}$,
\begin{equation*}
  E_{\theta_{-i}}[(u_{i}(f(\theta_{i},\theta_{-i}), \theta_{i}) - c_{i}(s^{*}_{i}(\theta_{i}), \theta_{i}))|\theta_{i}]
  \geq
  E_{\theta_{-i}}[(u_{i}(f(\hat{\theta}_{i},\theta_{-i}), \theta_{i}) - c_{i}(s^{*}_{i}(\hat{\theta}_{i}), \theta_{i}))|\theta_{i}].
\end{equation*}
 Note that the above inequality cannot infer the formula (\ref{old BIC}). Consequently, it cannot be inferred that there exists a direct mechanism that can truthfully implement $f$ in Bayesian Nash equilibrium. $\Box$

\textbf{Discussion 2:}
 Someone may disagree with Note 4 and Proposition 1, and propose a ``direct revelation game'' as follows. For a given social choice function $f$, suppose there is an indirect mechanism $\Gamma=(S_{1},\cdots,S_{I},g(\cdot))$ that implements $f$ in Bayesian Nash equilibrium, and the equilibrium strategy is $s^{*}=(s^{*}_{1}, \cdots, s^{*}_{I})$. Consider this equilibrium, there is a mapping from vectors of agents' types into outcomes. Now we take the mapping to be a revelation game, $i.e.$, each agent $i$ with private type $\theta_{i}$ independently chooses a type $\hat{\theta}_{i}\in\Theta_{i}$ to report to the designer, and the designer suggests each agent an action $s^{*}_{i}(\hat{\theta}_{i})\in S_{i}$. Then no type of any agent can benefit by reporting a false type $\hat{\theta_{i}}\neq\theta_{i}$ and performing the suggested action $s^{*}_{i}(\hat{\theta}_{i})$. As a result, truth-telling is the equilibrium strategy of this game, \emph{i.e.}, each agent $i$ reports his true type $\theta_{i}$ and performs the strategy action $s^{*}_{i}(\theta_{i})$, the same as what he would perform in the indirect mechanism $\Gamma$.

\textbf{Answer 2:}
It should be noted that in the direct revelation game, each agent $i$ with private type $\theta_{i}$ can choose any type $\hat{\theta}_{i}\in\Theta_{i}$ \emph{arbitrarily} to report to the designer, which means that the corresponding suggestion $s^{*}_{i}(\hat{\theta}_{i})\in S_{i}$ is not restricted to be $s^{*}_{i}(\theta_{i})$. Thus, after the designer receives a report profile $(\hat{\theta}_{1},\cdots,\hat{\theta}_{I})$, in order to exactly know which $s^{*}_{i}(\hat{\theta}_{i})$ should be suggested to each agent $i$, the designer must know each agent $i$'s strategy function $s^{*}_{i}(\cdot): \Theta_{i}\rightarrow S_{i}$ which is specified in an indirect mechanism $\Gamma=(S_{1},\cdots,S_{I},g(\cdot))$. However, in the framework of mechanism design, \emph{the designer is always at the information disadvantage in a mechanism}: she never knows each agent $i$'s private type $\theta_{i}$, nor his private strategy function $s^{*}_{i}(\cdot): \Theta_{i}\rightarrow S_{i}$. \footnote{Otherwise, assume to the contrary that the designer knows each agent $i$'s strategy function $s^{*}_{i}(\cdot): \Theta_{i}\rightarrow S_{i}$, then she can easily infer each agent $i$'s private type $\theta_{i}$ from his report $s^{*}_{i}(\theta_{i})$. Obviously, it contradicts the basic framework of mechanism design and does not hold.} Therefore, the so-called direct revelation game does not hold. $\Box$

\section {Conclusion}
This paper mainly investigates the notion of Bayesian Nash equilibrium of a mechanism when strategies of agents are costly actions. The work is also relevant to the possible failure of revelation principle. So far, there have been several discussions on possible failures of the revelation principle: Kephart and Conitzer \cite{Kephart2016} proposed that when reporting truthfully is costless and misreporting is costly, the revelation principle can fail to hold. Bester and Strausz \cite{Bester2001} pointed out that the revelation principle may fail because of imperfect commitment.

This paper proposes that when strategies of agents in an indirect mechanism are costly actions, the definition of Bayesian Nash equilibrium of the mechanism should be based on a profit function rather than the conventional utility function which does not include a parameter of agent's strategy. This is the key point why the revelation principle for Bayesian Nash equilibrium may fail, and this failure is different from the above-mentioned possible failures of the revelation principle.

\section*{Acknowledgments}
 The author is grateful to Fang Chen, Hanyue, Hanxing and Hanchen for their great support.

\end{document}